\documentstyle{article}

\title{\bf On the forward-backward correlations 
in a two-stage scenario}
\author{M.A.Braun$^1$, C.Pajares$^2$ and V.V.Vechernin$^1$\\
 $^1$ Department of High
Energy Physics,
University of St. Petersburg,\\ 198904 St. Petersburg, Russia\\ 
 $^2$ Departamento
de F\'{\i}sica de Part\'{\i}culas, Universidade de Santiago de
Compostela,\\ 15706-Santiago de Compostela, Spain}

\date{}
\def\beq{\begin{equation}}
\def\eeq{\end{equation}}

\def\avf {\langle F \rangle}
\def\avb {\langle B \rangle}

\def\df{\delta_{F,\sum F_i}}
\def\db{\delta_{B,\sum B_i}}
\def\pp{\prod_{i=1}^N p(F_i,B_i)}
\begin{document}
\maketitle
\vspace{1 cm}
\begin{center}
{\bf Abstract}
\end{center}
\vspace{1 cm}

It is demonstrated that in a two-stage scenario with elementary
Poissonian emitters of particles (colour strings) arbitrarily
distributed in their number and average multiplicities, the 
forward-backward correlations are completely determined by the final
distribution of  forward particles. The observed linear form of the
correlations then necessarily requires this distribution to have a
negative binomial form. For emitters with a negative binomial distribution
in the produced particles distributed so as to give the final distribution
also of the negative binomial form, the forward-backward correlations have
an essentially non-linear form which disagrees with the experimental data.  
\newpage
\section{Introduction}
Multiparticle production at high energy is currently (and successfully)
described in a two-stage scenario. At the first stage a certain
number of colour strings are formed, stretched between the incoming
partons. At the second stage these strings decay into the observed
secondary hadrons. In the simplest version the strings emit particles
independently. However one may also introduce the interaction between
strings in the form of their fusion and/or percolation. In any case the observed
multiplicity distribution is a convolution of the probability distribution
of the possible string configurations and the multiplicity distribution
generated by the individual strings. In fact both are not well-known, so that
to understand the dynamics of the multiparticle production one has to
reconstruct both from a single distribution of the observed secondaries.
To facilitate this task it is natural to use some other experimental
information about the multiparticle production. An immediate candidate is
the long-range correlation data, currently studied as the forward-backward
correlations (FBC). They correspond to studying the average multiplicity 
in, say, the
backward hemisphere $\avb$ as a function of the event multiplicity in 
the forward hemisphere $F$. 

The most striking feature of the data is that they can be
almost perfectly represented by a linear function [1-4]
\beq
\avb=a+bF,
\eeq
the strength of the correlation measured by the coefficient $b$.
The dynamical reason for this simple behaviour is not known.
In this paper we try to understand this behaviour in the two-stage scenario,
decribed above.

Our starting point is the assumption that individual emitters 
(colour strings) have a Poisson
distribution in the number of produced particles. This assumption is based on
the idea that a string is an homogeneous extended object in the rapidity space.
Its fundamental property is that parts of the string
occupying different regions in the rapidity space do not interfere (except
at small rapidity distances). Then the particle distributions coming from 
parts of
the string should have basically the same form which should also be similar to
the overall particle distribution (except for the average multiplicities, which
shoud sum together into a total one). In terms of the corresponding generating
functions
(GF) this means that the functional dependence of the product of GF
corresponding to different parts of the string should be the same and coincide
with the functional dependence of the total GF for the string. As we shall
see, this condition  is
fulfilled by the exponential dependence of the Poisson GF (see Eq.
(22)). It is difficult to imagine any other probability distribution with the
same properties.

Our main result is that with the individual emitters
having the Poisson multiplicity distribution, the FBC are completely
determined by the final multiplicity distribution in the forward rapidity window
$P(F)$ and do not depend explicitly on the distribution of strings themselves 
(and consequently on their dynamics).
If, in accordance with the experimental data, one further requires that 
$\avb$ be a linear function of $F$, then one finds that the corresponding
$P(F)$ must be a negative binomial
distribution (NBD), also in agreement with the experimental  data.
Thus our explanation of the linear dependence (1) is that it comes from a
Poissonian form of the elemental distribution, which the distribution in the
elemental emitters transforms into a NBD. As we shall see such transformation
is only possible with different kinds of strings, including those with very
small multiplicities. Such strings are indeed present in current scenarios,
either as "short" strings, extended over comparatively small rapidity intervals,
in the independent string picture, or strings with small transverse area, which
appear in the percolation scenario.

This result is not at all trivial, as one can conclude from an alternative
picture in  which both the strings and the final distribution  are negative
binomial.  In this case we find that $\avb$ is not a linear function on $F$, but
rather a function which grows with $F$ with a falling derivative, so that at
high $F$ it either saturates or grows only logarithmically.

To compare our predictions with (scarce) experimental data we take into
account that most (if not all) particles are created as resonances. So in fact
our numbers $F$ and $B$ refer not to the observed particles but only to
the directly formed ones. As in most other approaches we then have to introduce
an average number $r$ of the observed particles per single directly produced
one (resonance, or cluster). The number of observed particles in the
forward or backward rapidity windows will then be $n_F=rF$ and $n_B=rB$
respectively. Of course, from (1) also a linear relation will follow
between $\langle n_B\rangle$ and $n_F$ with the same parameter $b$.
Using the experimental data on the forward multiplicity distribution
in [5] and comparing them with the data on the parameter $b$ [4]
we find that  $r$ does not seem to depend on energy and its value is around  3.

\section{Formalism}

Let us consider a set of $N$ identical elemental emitters
distributed with a probability $w(N)$.
We denote $p(F,B)$ the probability to find $F(B)$ particles in the forward
(backward) hemispheres from a single emitter.
The overall distribution in $F,B$ coming from any number of emitters is
given by
\beq
P(F,B)=\sum_Nw(N)\sum_{F_i,B_i}\df\db\pp.
\eeq
The overall averages are
\beq
\langle A\rangle=\sum_{F,B}A(F,B)P(F,B).
\eeq
The conditional probability to find $B$ particles in the backward hemisphere
with their number $F$ in the forward hemisphere fixed is
\beq
P_F(B)=\frac{P(F,B)}{P(F)}.
\eeq
Here $P(F)$ is the (unconditional) probability to observe $F$ particles in the
forward hemisphere
\beq
P(F)=\sum_{B}P(F,B).
\eeq
According to (4) and (5)
the average with a given $F$ is given by the ratio
\beq
\langle A\rangle_F=\frac{\sum_{B}A(F,B)P(F,B)}{\sum_{B}P(F,B)}.
\eeq

It is convenient to pass to the formalism of 
generating functions (GF).
One  represents the Kronecker symbols entering (2) as
\beq
\delta_{m,n}=\int\frac{dx}{2\pi i x}x^{n-m},
\eeq
where the integration contour in the complex $x$ plane goes around the
origin. Using this formula one writes for the probability (5)
\beq
P(F,B)=\int\frac{dx}{2\pi i x}x^{-F}\int\frac{dy}{2\pi i y}y^{-B}
\sum_Nw(N)\prod_{i=1}^N\Big(\sum_{F_i,B_i}x^{F_i}y^{B_i}p(F_i,B_i)\Big).
\eeq
Each sum inside the product over $i=1,...N$ gives a GF for
the distribution $p(F,B)$ from a single emitter:
\beq
g(x,y)=\sum_{F,B}x^{F}y^{B}p(F,B).
\eeq
So we get
\beq
P(F,B)=\int\frac{dx}{2\pi i x}x^{-F}\int\frac{dy}{2\pi i y}y^{-B},
 G(x,y)
\eeq
where we  introduced the overall GF
\beq
G(x,y)=\sum_Nw(N)g^N(x,y).
\eeq

The averages with a fixed $F$ can be readily expressed in terms of $G$.
We find
\beq
\sum_BA(F,B)P(F,B)=
\int\frac{dx}{2\pi i x}x^{-F}\int\frac{dy}{2\pi i y}\sum_BA(F,B)y^{-B}
G(x,y).
\eeq
To do the sum over $B$ we use
\beq
\sum_BA(F,B)y^{-B}=
A\left(F,-y\frac{\partial}{\partial y}\right)\sum_By^{-B}=
A\left(F,-y\frac{\partial}{\partial y}\right)\frac{y}{y-1}.
\eeq
The integration  by parts in $y$ then gives
\beq
\sum_BA(F,B)P(F,B)=
\int\frac{dx}{2\pi i x}x^{-F}\int\frac{dy}{2\pi (y-1)}
A\left(F,y\frac{\partial}{\partial y}\right)
G(x,y).
\eeq
Taking finally the residue at $y=1$ we find
\beq
\sum_BA(F,B)P(F,B)=
\int\frac{dx}{2\pi i x}x^{-F}
\Big[A\left(F,y\frac{\partial}{\partial y}\right)
G(x,y)\Big]_{y=1}.
\eeq

In partcular we get
\beq
P(F)=\sum_BP(F,B)=\int\frac{dx}{2\pi i x}x^{-F}G(x,1)
\eeq
and
\beq
\sum_BBP(F,B)=\int\frac{dx}{2\pi i x}x^{-F}G^{\prime}_y(x,1).
\eeq
This allows to write the average number of particles in the
backward hemisphere, given their number in the forward hemisphere,
as
\beq
\langle B\rangle_F=\frac
{\int dx x^{-F-1}G^{\prime}_y(x,1)}
{\int dx x^{-F-1}G(x,1)}.
\eeq

This formalism can be generalized in a straightforward manner to the case when
the emitters are not identical. Call a configuration $C$ a particular
number $N_C$ of emitters enumerated by $i=1,2,....N_C$ all of them (or part
of them) different. Let the probability to find $F(B)$ particles in the
forward (backward) hemisphere from the $i$-th emitter be $p_i(F,B)$.
Define corresponding GF $g_i(x,y)$  as in (9).
Then for the overall GF one finds an evident generalization of (11)
\beq
G(x,y)=\sum_{C}w(C)\prod_{i=1}^{N_C}g_i(x,y),
\eeq
where $w(C)$ is the probability to find a given configuration $C$ and 
summation goes 
over all possible configurations.

\section{Poisson emitters}

In the following we assume that individual emitters have no FBC, which implies
that both the probability and GF for them factorize
\beq
p(F,B)=p(F)p(B),\ \ g(x,y)=g(x)g(y).
\eeq
As to the concrete form of the distributions $p(F)$ and $p(B)$, we assume them
identical and Poissonian:
\beq
p(F)=e^{-\alpha}\frac{\alpha^F}{F!}.
\eeq
The corresponding GF is an exponential:
\beq
g(x,y)=g(x)g(y),\ \  g(x)=e^{\alpha(x-1)}.
\eeq
The reason to assume the Poissonian
distribution for a single emitter has been explained in the Introduction:
it naturally goes with the notion of the string as a homogeneous extended
object in the
rapidity space.

Passing to the overall GF $G$, we see that
it does not generally factorize. With (22) we find
\beq
G(x,y)=\sum_Nw(N)e^{Na(x+y-2)}.
\eeq
The only case when it does factorize corresponds to a fixed number of
emitters $w(N)=\delta_{N,N_0}$.

The special property of the Poisson distribution for the elemental emitters
shows itself in  that the overall GF results depending only on the sum
$x+y$. As a consequence we find
\beq
G^{\prime}_y(x,1)=G^{\prime}_x(x,1).
\eeq
Putting this into (17) and
integrating  by parts in $x$ we transform the numerator in (18) as
\beq
(F+1)\int\frac{dx}{2\pi i x}x^{-F-1}G(x,1)=(F+1)P(F+1).
\eeq
As a result $\avb_F$ becomes expressed entirely in terms of the distribution
in the number of the forward particles:
\beq
\avb_F=\frac{(F+1)P(F+1)}{P(F)}.
\eeq

Before we go further, note that this result holds not only for identical
emitters, but also for  different but
Poissonian emitters. In fact the key point is that for an individual
Poissonian emitter
\beq
p_i(x,y)=e^{\alpha_i(x+y-2)}=p_i(x+y)
\eeq
depends only on the sum $x+y$. Summation over configurations in (19) will not
change this result, so that the crucial identity (24) and the following
equations will be valid.

Eq. (26) allows to find $\avb$ once the particle distribution $P(F)$ in the
forward rapidity window is known. A more detailed form of Eq. (26) can be
obtained in the approximation of the KNO scaling, when $\avb$ can be
expressed through  the
scaling function (see Appendix 1.).

\section{Linear form of $\avb_F$}

As we have shown, under the assumption that elemental emitters are Poissonian
$\avb_F$ can be expressed entirely in terms of the distribution
in the number of the forward particles by Eq. (26).
Let us assume that the dependence of $\avb_F$ is linear and given by (1).
Then from (26) we obtain a recurrent relation
\beq
P(F+1)=\frac{a+bF}{F+1}P(F)=\left(b+\frac{c}{F+1}\right)P(F),
\eeq
where
$ c=a-b$.
As we demonstrate below, it allows to completely determine $P(F)$
(see also [6]).

We put
\beq
P(F)=b^FQ(F).
\eeq
Then we find a recurrency for $Q(F)$
\beq
Q(F+1)=\left(1+\frac{\lambda}{F+1}\right)Q(F).
\label{Qrec}
\eeq
where
$\lambda=c/b=a/b-1.$
Recurrency (\ref{Qrec}) is trivially solved by
\beq
\ln Q(F)=\ln Q(0)+\sum_{n=1}^{F}\ln\left(1+\frac{\lambda}{n}\right).
\eeq

The sum over $n$ can be expressed via the Gamma-function.
Indeed we find that if
\beq
S(\lambda)\equiv \sum_{n=1}^{F}\ln\left(1+\frac{\lambda}{n}\right)
\eeq
then its derivative in $\lambda$
\beq
S^{\prime}(\lambda)=\sum_{n=1}^{F}\frac{1}{n+\lambda}=
\psi(F+\lambda+1)-\psi(\lambda+1).
\eeq
Integrating this over $\lambda$ in the interval $[0,\lambda]$ we get
\beq
S=\ln\frac{\Gamma(F+\lambda+1)}{\Gamma(F+1)\Gamma(\lambda+1)}
\eeq
This finally gives for the distribution $P(F)$:
\beq
P(F)=P(0)e^{F\ln b}\frac{\Gamma(F+\lambda+1)}{\Gamma(F+1)\Gamma(\lambda+1)},
\label{PFg/g}
\eeq
where $P(0)$ is to be determined from the normalization condition.
Evidently one should have $b<1$ for normalizability.

One observes that (\ref{PFg/g})
is nothing but a negative binomial distribution
with the average value $\langle F\rangle$ and parameter $k$ determined by
\beq
b=\frac{\langle F\rangle}{\langle F\rangle +k},\ \ k=\lambda+1=\frac{a}{b}.
\eeq
Thus the NBD is singled out from all other distributions by the fact that if
the elemental emitters are Poissonian it leads to a linear dependence of
$\avb$ on $F$. However to give this result a physical meaning we still have to
demonstrate that it is possible to produce a NBD distribution  by a
convolution of Poisson distributions with a physically admissible
distribution of emitters. This problem is discussed in the next section.

\section{Negative binomial distribution from Poisson distributions}
We start by showing that if all the emitters are identical and Poissonian,
one cannot produce a NBD distribution of final particles with any physically
admissible distribution of emitters.
Indeed let $P(F)$ have a NBD form.
The corresponding GF is then
\beq
G_{NBD}(z)=(a-bz)^{-k},\ \ a-b=1.
\eeq
The distribution itself can be extracted by developing this function
in a power series:
\beq
P_{NBD}(F)=a^{-k}C_{k-1+F}^{F}\left(\frac{b}{a}\right)^F,
\eeq
where we define the binomial coefficient as
\beq
C_{k-1+F}^{F}=\frac{\Gamma(F+k)}{\Gamma(k)\Gamma (F+1)}
\eeq
The average value of the multiplicity
is given by
\beq
\langle F\rangle=k(a-1),
\eeq
so that the NBD is fully determined by two parameters: the average
multiplicity and $k$.

Now it is trivial to show that if the GF are known for
both the individual emitter and for the overall distribution, the
GF for the distribution of emitters is determined 
uniquely (provided all emitters are identical). Indeed, considering again the
distribution of only forward particles, we have seen that the overall GF is
given by
\beq
G(z)=\sum_Nw(N)g^N(z),
\eeq
where $g(z)$ is the GF for the individual emitter. Introducing the GF
for the distribution of emitters by
\beq
H(z)=\sum_Nw(N)z^N,
\eeq
we observe that
\beq
G(z)=H(g(z)).
\eeq
Assume that we know both $G(z)$ and $g(z)$. We solve $g=g(z)$ for $z$
to determine $z=z(g)$. Putting this into $G(z)$ we find the unknown 
function $H(z)$:
\beq
H(g)=G(z(g)).
\eeq
As we see, the GF for the emitter distribution is determined in a unique manner.
However it may be or may not be physical, depending on its behaviour at small
values of argument. For a physically valid $H(z)$ it should be analytic
at small $z$ with all the derivatives positive. Otherwise it cannot be
interpreted as a GF for a physically sensible distribution.

Let us see what this means for the case when the final distribution is
negative binomial and the individual emitter is Poissonian.
Then the GF for an individual emitter is given by (22)
Inverting  we find
\beq
z=1+\frac{1}{\alpha}\ln g.
\eeq
So with the final GF given by (37), the GF for the emitter distribution
turns out to be
\beq
H(g)=\Big[a-b\left(1+\frac{1}{\alpha}\ln g\right)\Big]^{-k}=
\Big[1-b\frac{1}{\alpha}\ln g\Big]^{-k}.
\eeq
This function is not analytic at $g=0$, all its derivatives being infinite
at the origin. So it cannot be interpreted as a GF for some physically
sensible distribution of emitters.

Thus there does not exist a distribution of identical Poissonian emitters which
gives an overall NBD. However this is not true if the emitters are allowed
to be different and, in particular, continuously distributed in their
average values $\alpha_i$.
Indeed, for different Poissonian emitters, we have for the GF for the
forward hemisphere
\beq
G(z)=\sum_Cw(C)\prod_{i=1}^{N_C}e^{\alpha_i(z-1)}=\sum_Cw(c)e^{\alpha(C)(z-1)},
\eeq
where
\beq
\alpha(C)=\sum_{i=1}^{N_C}\alpha_i
\eeq
and has a meaning of the average forward multiplicity for a given
configuration $C$.
The quantity to be averaged over configurations in (47) depends only on 
$\alpha(C)$. Assuming it to vary continuously, one can rewrite (47) as
\beq
G(z)=\int_0^{\infty} d\alpha w(\alpha)e^{\alpha(z-1)},
\eeq
where $w(\alpha)>0$ now has the meaning of the probability distribution for
the average forward multiplicity of the emitters, normalized
accordingly
\beq
\int_0^{\infty} d\alpha w(\alpha)=1.
\eeq

Now it is trivial to find a form of $w(\alpha)$ which leads to the NBD.
In fact let us take
\beq
w(\alpha)=\frac{r^k}{\Gamma(k)}\alpha^{k-1} e^{r\alpha}.
\eeq
Then (47) gives
\beq
G(z)=\left(\frac{r}{r+1-z}\right)^k=(a-bz)^{-k}, 
\eeq
where
\[a=1+1/r,\ \ b=1/r.\]
So we have indeed obtaind the NBD (37). From the derivation it can be seen that
the crucial point in obtaining a GF which behaves as a meromorphic
function, instead of the entire Poissonian, has been the integration over
small valuses of $\alpha$ down to $\alpha=0$. Physically this implies existence
of emitters (strings) with very small average multiplicities.

Note that a construction similar to (49) with (51) was used in the VENUS model
[7] to obtain a realistic NBD for p-A spectra. 

To conclude, in the picture where the Poissonian emitters are different
and may have very small average multiplicities, it is possible to choose their
distribution to have the overall NBD. In this picture the FBC have an exact
linear dependence as in the empirical formula (1) with the parameter $b$
determined by (36) 

\section{Negative binomial distribution from negative binomial
distributions}
Although the results of the preceding section show that one can assume the
elemental emitters to be Poissonian and end up with the NBD for the finally
produced particles, provided the emitters are different and extend to
very small average multiplicities, it presents some interest to see 
what happens in a more conservative picture, in which all elemental
emitters are identical.
As we have seen, to have the final NBD the emitters
cannot be Poissonian in this case. So the first problem is to find an
adequate distribution for a single emitter which would lead to the NBD with an
approprate physically meaninful emitter distribution. 
This problem can evidently have many solutions. One of the simplest was found 
by C.Iso and K.Mori [8]. In the following we use their results to study
the long-range correlations which follow from their form of the
distributions. We expect that the conclusion will not depend too heavily
on this particular choice. In any case they will serve as a simple example
to be compared with the Poissonian emitters.
The solution of C.Iso and K.Mori is to take the GF for the individual
emitter essentially also of the NBD form
\beq
g(z)=\frac{e}{d}-\frac{1}{d}(a-bz)^{-y},
\eeq
with the requirements:
\beq
y>0, \ \ d=e-1<0,\ \ e=a^{-y}.
\eeq
The latter condition follows from the requirement that the emitter should emit
at least one particle, i.e. $g(0)=0$. The average multiplicity from a single
emitter corresponding to (53) is given by
$
\langle F\rangle/xd
$
,where $\langle F\rangle$ is the overall average corresponding to the NBD.
With (53) one easily shows that the GF for the emitter distribution has a form
\beq
H(z)=(e-dz)^{-x},
\eeq
where $x<0$ is determined by the final BND parameter $k$:
$k=-xy$.
It turns out that (55) corresponds to a physically meaningful distribution
so long as
$
-x\geq 2
$.
The solution of C.Iso and K.Mori contains only one more parameter $y$ (or $x$)
in addition to the two parameters of the final NBD.

Passing to the FBC, with a non-Poissonian emitters 
our simple formula (26) 
does not hold and we have to explicitly calculate them from (18).
We have to calculate
\beq
{G}^\prime_u(z,1)=\Big[\sum w(N)g^N(z)Ng^{N-1}(u){h}^\prime(u)\Big]_{u=1}.
\eeq
Using the explicit form of $g(u)$, Eq. (57) we find
\beq
g^\prime(1)=\Big[-\frac{by}{d}(a-bu)^{-y-1}\Big]_{u=1}=-\frac{by}{d}.
\eeq
We are left with
\beq
{G}^\prime_u(z,1)=-\frac{by}{d}\sum w(N)Ng^N(z)=
-\frac{by}{d}g\frac{d}{dg}\sum w(N)g^N=
-\frac{by}{d}g\frac{d}{dg}(e-dz)^{-x}.
\eeq
In the last equation we used (55). Taking the derivatives and putting (37) for
$g(z)$ we finally obtain
\beq
{G}^\prime_u(z,1)=\frac{bk}{d}\Big[e(a-bz)^{-k+y}-(a-bz)^{-k}\Big].
\eeq
Both terms are the GF for the NBD, with parameters $k-y$ and $k$ respectively.
So projecting their $F$-th term in the development in the power series in $z$ 
will give
\[eP_{NBD}(F,k-y)-P_{NBD}(F,k).\]
According to (18) to obtain $\avb$ we have to divide this by $P_{NBD}(F,k)$.
This brings us to our final result
\beq
\avb=\frac{bk}{d}\Big[\frac{P_{NBD}(F,k-y)}{P_{NBD}(F,k)}-1\Big],
\eeq
or using relations between $e$, $d$ and the overall average $\langle F\rangle$
\beq
\avb=\frac{\langle
F\rangle}{1-a^{-y}}\Big[1-a^{-y}\frac{P_{NBD}(F,k-y)}{P_{NBD}(F,k)}\Big],
\eeq
where 
\[a=\frac{\langle F\rangle}{k}+1\]
and $y$ is the only extra parameter introduced in the solution.
Using the explicit expression for the NBD distribution we find
\beq
\avb=\frac{\langle
F\rangle}{1-a^{-y}}\Big[1-\frac{\Gamma(k)}{\Gamma(k-y)}
\frac{\Gamma(F+k-y)}{\Gamma(F+k)}\Big].
\eeq

At large values of $F$ the negative second term is vanishing as 
$(F+k)^{-y}$. So for $y>0$ $\avb$ grows with $F$ tending to a constant value,
which is somewhat greater than its overall average. In the limiting case 
$y=0$  one gets
\beq
\avb=\frac{\langle
F\rangle}{a}\Big[\psi(F+k)-\psi(k)\Big].
\eeq
So $\avb$ continues to grow indefinitely with $F$, but only logarithmicallly.
Thus, comparing with the
Poisson emitters case, we find the same tendency in the behaviour of $\avb$: 
it grows with $F$. However now the growth is either stopped at high $F$ or
severely moderated. 

\section{Comparison to the experiment. Discussion.}
Passing to the comparison with the latest experimental data [4], we first have 
to
stress that the observed values of $b$ strongly depend on the rapidity windows
chosen. They diminish substantially with the gap between the
forward and backward windows. This is thought to be an effect of the remaining
short range correlations. However the fact that $b$ continues to considerably go
down when the gap is raised from 4.37 to 5.43 makes this explanation dubious.
In any case there is little reason to think that the FBC are universal for
different choices of the rapidity windows, so that the analysis has to be
made for each particular choice separately. To get rid of the short range
correlations the gap should be taken large enough. From this point of view
the data which involve the full forward and backward spaces are evidently
not suitable.
On the other hand our results are presented in terms of the particle
distribution in the forward window. Such data are rather scarce in general and
still more scarce if we want to have them exactly for the windows which have
been used for the determination of the FBC.
Thus, comparing the data from [4] and [5] we find only  one 
pseudorapidity window
\beq 2.18<|\eta|<3.25\eeq
in which the parameter $b$ has been measured  [4] and which is covered by
two pseudorapidity windows of nearly the same length
\beq 2.5<|\eta|<3.5,\ \ {\rm and}\ \ 2.0<|\eta|<3.0\eeq
where the particle distribution has been studied  [5].
The  data on the FBC  in the window agree with the form (1) with 
the parameter $b$ 
given in the Table 1 for various c.m, energies. As to the particle distributions
in the two rapidity windows (65), it is claimed to be well described bu the NBD
with parameters listed in Table 2 for two c.m. energies 200 and 900 GeV.
As we see neither the windows nor the energies do not match in the two
sets of data. In the following we use the values of the NBD parameters for the
interval (64) as the average for the two intervals (65) and also use the
parameters in Table 2 for energies 200 and 900 to desribe $b$ at 300 and 1000 GeV
respectively.

Turning to our results we first stress that the form of $\avb$ which follows
from the NBD for individual strings (62) or (63) is in striking contradiction
with the data. In Figure  we illustrate this behaviour for two possible values
of the parameter $y=0$ and $0.5$ and compare it with the straight line
corresponding to $b=0.252$ and $a=1.2$ ([4], with an appropriate
reduction of $a$ for the interval (64)). The number
$r$ of observed particles per directly produced was taken unity. However
changing it does not improve the situation at all. So our first conclusion is
that the scenario of strings with the NBD, which produce the overall NBD, does
not work.

In our basic picture with Poissonian strings the linear form (1) holds exactly.
The value of the parameter $b$ is given by Eq. (36) where the average $\langle
F\rangle$ refers to the directly produced particles. Passing to the observed
particles we find
\beq
b=\frac{\langle n_F\rangle}{\langle n_F\rangle+rk},
\eeq
where $r$ is the number of observed particle per one directly produced.
The data shown in Tables 1 and 2 allow to extract the value of $r$ for the
rapidity window (64) and two c.m. energies 200 and 900 GeV. We  find
\[
r=3.22\pm 0.68,\ \ \sqrt{s}=200\ {\rm GeV},\]
\beq
r=2.63\pm 0.52,\ \ \sqrt{s}=900\ {\rm GeV}.\eeq
Note the large errors, which come from the errors in the experimental $b$ and
fitted
$k$. From these values it is difficult to extract any definite information as
to the behaviour of $r$ with energy. They are compatible with $r$ independent
of energy altogether.

If we forget about the necessity to cut the central rapidity window and
consider the whole phase space both in the forward and backward hemispheres,
we get values of $r$ with less errors. They are presented in the first
column of Table 3 for various energies. They also seem to be independent of
energy.

Turning to the literature, we find that the FBC have been mostly discussed
for the whole phase space  in
terms of the assumption by Chou and Yang [9] about the form of the
multiplicity distributions in the total number of particles $n$
and in the difference
$z=n_F-n_B$. On purely phenomenologal grounds Chou and Yang assumed that this
distribution is a product of the NBD in $n$ and a binomial in $z$, the latter
proportional to 
 $C_{n/r}^{n_F/r}$,
where $r$ is the number of particles per cluster. 
From this assumption,
it
was found in [10] by numerical calculation that the resulting behaviour of $\avb$
as a function of
$F$ was close to linear. In this approach
the parameter $b$ can be expressed through the total multiplicity as
\beq
b=\frac{\langle n\rangle+k-r}{\langle n\rangle+k+r}.
\eeq
Comparing this formula with the experimentally observed NBD parameters
the values of $r$ were found  for various energies in [4]. 
They are presented in the second column of Table 3. As we see they are smaller
than ours and seem to grow with energy.

Commenting on these results we have first to mention that our scenario with
Poissonian strings automatically leads to the distribution in the difference
$z=F-B$ of the binomial type (see Appendix 2.). So in this way the
formal assumption of Chou-Yang receives a concrete realization in our scheme.
Since the linear dependence (1) is exact in our approach, we have little
doubt that it should also hold in [10], which the numerical calculations
made there confirm.
However there remain  a quantitative difference in the two approaches.
One
observes that for $r$ different from unity the final formulas for $b$
(36) and (68)  remain  different if one considers the whole phase space and
takes into account that
$\langle n\rangle=2\langle F\rangle$. Only for $r=1$ they coincide.
As a result,
our values for $r$
are systematically larger than those obtained in [4] from the essentially
Chou-Yang approach 
(cf. the two columns in the Table 3). On the other hands
our values do not essentially change with energy, whereas those of [4] grow, so
that the difference between them seems to be disappearing at larger energies.
The exact origin of this difference is easy to trace.
It is related to the way in which the
distribution in clusters is transformed into the one for the observed
particles.  In both approaches this transformation is made in an approximate
manner. In the purely phenomenological  approach of Chou-Yang the distribution
in the total number of observed particles is taken to be the NBD just on
experimental grounds. However it does not evidently correspond to the idea that
particles are formed from clusters. In fact if we assume that all particles are
formed from clusters which decay in exactly $r$ particles, their  distribution
cannot be an NBD in $n$, but rather an NBD in $n/r$. In this case one will only
find a non-zero probability to see $r, 2r, 3r ...$ observed particles. In our
approach the distribution in $n$ is taken to be an NBD in $n/r$ suitably
continued to non-integer values of the argument. This is evidently also only an
approximation. It is trivial to find the exact distribution in $n$ which
corresponds to cluster formation and decay in our approach (see Appendix 2.).
Understandably it is not an NBD in the general case, passing into it only when
the number of particles into which a cluster may go is fixed.

In conclusion we think that due to  the introduction of a certain
dynamical mechanism for particle production via interacting strings
our approach allows not only to understand the reason for the formal
assumption made in [9] and exploited in [4,10], but also presents some
possibilities to improve the description of the FBC. In particular
concrete models for string fusion and percolation introduce well
defined distributions of strings and so allow, in principle, to predict
the final form of the observed particle distribution. Unfortunately
quantitative realization of this idea necessarily requires taking into account
energy-momentum conservation at each step, which can only be achieved 
numerically, in a well developed Monte Carlo algorithm.

\section{Acknowledgements}
This work has been done under the contract AEN 99-0589-C02-02
from CICYT of Spain. 

\section{Appendix 1. KNO scaling functions}
Using the KNO scaling for $P(F)$  we present it as
\beq
P(F)=\avf^{-1}\psi\left(\frac{F}{\avf}\right)
\eeq
with some smooth scaling function $\psi(x)$. Then
\beq
P(F+1)=\avf^{-1}\psi\left(\frac{F+1}{\avf}\right)=
P(F)+\avf^{-2}\psi^{\prime}\left(\frac{F}{\avf}\right).
\eeq
Putting this into (26) we obtain 
\beq
\avb_F=F+1+\frac{F+1}{\avf}[\ln\psi(x)]^{\prime}_{x=F/\avf}.
\eeq
This formula is generally quite different from (1). Its exact form depends on
the form of the KNO scaling function. Let us assume
\beq
\psi(x)=Ae^{-\beta|x-1|^{\alpha}}.
\eeq
Then
\beq
[\ln\psi(x)]^{\prime}=-{\rm sign}(x-1)\alpha\beta|x-1|^{\alpha-1},
\eeq
so that we finally get
\beq
\avb_F=F+1+{\rm sign}(F-\avf)\alpha\beta
\Big(\Big|\frac{F}{\avf}\Big|^{\alpha-1}-
\Big|\frac{F}{\avf}\Big|^{\alpha}\Big).
\eeq
As one observes, as a rule, the $F$ dependence is much more compliciated than
the linear one assumed in (1).

\section{Appendix 2. The distribution in $z$ and production of clusters}
We first show that the Poisson distribution of particles produced from the
individual emitter automatically leads to a binomial distribution in $z$ at
fixed total number of produced particles. (It is trivial to show that also the
inverse is true: the binomial distribution in $z$ means that the elemental
emitters are Poissonian). We again use the formalism of GF. In this language
transition from numbers of particles $F$ and $B$ to $n=F+B$ and $z=F-b$
corresponds to the transition from variables $x$ and $y$ to $u$ and $v$
determined by
\beq
x=uv,\ \ y=u/v.
\eeq
The GF $G(u,v)$ is related to the distribution $P(n,z)$ via
\beq
G(u,v)=\sum_{n=0}\sum_{z=-n}^{n}u^nv^z.
\eeq
Now assume that the elementary emitters are Poissonian so that $G(x,y)=G(x+y)$.
We present it as
\[
G(x+y)=\sum_na(n)(x+y)^N=\sum_na(n)(u/2)^n(v+1/v)^n=
\sum_na(n)(u/2)^n\sum_{m=0}^nC_n^mv^{n-2m}\]\beq=
\sum_na(n)(u/2)^n\sum_{z=-n}^nC_n^{(n+z)/2}v^{z}.
\eeq
From this expansion we conclude that
\beq
P(n,z)=(1/2)^na(n)C_n^{(n+z)/2}=(1/2)^n a(n)C_n^F
\eeq
and the overall distribution in $n$ is
\beq
P(n)=\sum_{z=-n}^nP(n,z)=a(n).
\eeq
Dividing (78) by (79) we get the distribution in $z$ for fixed $n$ , which
results proportional to $C_n^F$.

Now we discuss the correct treatment of resonance formation. Let the elemental
emitter (string) produce two types of resonances 1 and 2 with a probability
$p(n_1,n_2)$. Let the resonances decay into $r_1$ and $r_2$ observed particles
respectively. The final distribution of the observed particles will be given
by an evident generalization of Eq. (2):
\beq
P(n)=\sum_Nw(N)\sum_{n_{1i},n{2_i}}\delta_{n-\sum_i(r_1n_{1i}+r_2n_{2i})}.
\prod_ip(n_{1i},n_{2i})
\eeq
The  GF for the distribution in $n$ will be given by
\beq
G(u)=\sum_nP(n)u^n=\sum_Nw(N)
\Big(\sum_{n_1,n_2}p(n_1,n_2)u^{r_1n_1+r_2n_2}\Big)^N.
\eeq
If we introduce the GF for a single emitter
\beq
g(u_1,u_2)=\sum_{n_1,n_2}p(n_1,n_2)u_1^{n_1}u_2^{n_2},
\eeq
then Eq. (81) can be rewritten as
\beq
G(u)=\sum w(N)g^N(u^{r_1},u^{r_2}).
\eeq

Now assume that the two resonances are produced independently and the
distribution in both is Poissonian:
\beq
g(u_1,u_2)=e^{\alpha_1(u_1-1)+\alpha_2(u_2-1)},
\eeq 
where $\alpha_{1,2}$ give the average number of each sort of resonance produced
by a single emitter. Also assume that the emitters are distributed like
in (49) with the ratio $\alpha_2/\alpha_1$ fixed.
Then in complete analogy with (52) we shall get an overall GF in the
form
\beq
G(u)=\Big(a-b_1u^{r_1}-b_2u^{r_2}\Big)^{-k}
\eeq
with $a-b_1-b_2=1$.
Evidently this is not a NBD. It passes into a NBD only in two limiting cases:
$b_2=0$ when it gives a NBD in the observed particles and $b_1=0$ when it is
a NBD in resonances and $n$ takes on values which are multiples of $r$. 

\section{References}

\hspace{0.6cm}

1. S.Uhlig et al, Nucl. Phys. {\bf B132} (1978) 15.

2. UA5 coll., K.Alpgard et al., Phys. Lett., {\bf B 123} (1983) 361.

3. UA5 coll., R.E.Ansorge et al.,Z.Phys. {\bf C 37} (1988) 191.

4. E735 coll., T.Alexopulos et al., Phys. Lett. {\bf B 353} (1995) 155.

5. UA5 coll., R.E.Ansorge et al., Z.Phys. {\bf C 43} (1989) 357.

6. A.Giovanini and L.Van Hove, Z.Phys. {\bf 30} (1986) 242.

7. K.Werner and M.Kutschera, Phys.Lett. {\bf B 220} (1989) 243;  
K.Werner, Phys.Rep. {\bf 232} (1993) 87.
 
8. C.Iso and K.Mori Z.Phys. {\bf C46}(1990) 59.

9. T.T.Chou and C.N.Yang, Phys. Lett. {\bf 135} (1984) 175.

10. S.L.Lim, Y.K.Lim, C.H.Oh and K.K.Phua, Z.Phys. {\bf C 43} (1989) 621.

\section{Figure caption}
$\avb$ as a function of $F$ with the NBD for individual emitters.
The lower curve corresponds to $y=0.5$ in Eq. (62), the upper one
corresponds to $y=0$ (Eq. (63)). The straight line shows the experimental
behaviour at c.m.energy 200 GeV. 

\section{Tables}
\begin{center}
{\bf Table 1. Experimental data on $b$ for the rapidity window
(64) [4]}\vspace{0.5 cm}

\begin{tabular}{|l|r|r|r|r|}\hline
Energy (GeV)&300&546&1000&1800\\\hline
b&$0.252\pm 0.044$&$0.376\pm 0.033$&$0.381\pm 0.0021$&$0.501\pm 0.008$\\\hline
\end{tabular}
\end{center}
\vspace*{1 cm}
\begin{center}
{\bf Table 2. The NBD parameters for the rapdity windows (65) [5]}
\vspace{0.5 cm}

\begin{tabular}{|c|l|r|r|}\hline
Energy (GeV)&$\eta$-interval&$\langle F\rangle$& k\\\hline
200&$2.0-3.0$&$2.40\pm 0.03$&$2.01\pm 0.07\pm 0.10$\\\hline
200 &$2.5-3.5$&$2.14\pm 0.03$&$2.2\pm 0.1\pm 0.1$\\\hline
900&$2.0-3.0$&$3.72\pm 0.05$&$2.2\pm 0.1\pm 0.1$\\\hline
900&$2.5-3.5$&$3.41\pm 0.05$&$2.2\pm 0.1\pm 0.1$\\\hline
\end{tabular}
\end{center}
\vspace*{1 cm}
\begin{center}
{\bf Table 3. Number $r$ of particles per produced cluster}
\vspace{0.5 cm}

\begin{tabular}{|c|r|r|}\hline
Energy (GeV)& r, Eq. (66)& r, Eq (68) [4] \\\hline
300 &$2.86\pm 0.29$&$2.27\pm 0.24$\\\hline
546 &$2.73\pm 0.26$&$2.31\pm 0.19$\\\hline
1000&$3.11\pm 0.30$&$2.65\pm 0.25$\\\hline
1800&$2.96\pm 0.21$&$2.64\pm 0.16$\\\hline
\end{tabular}
\end{center}

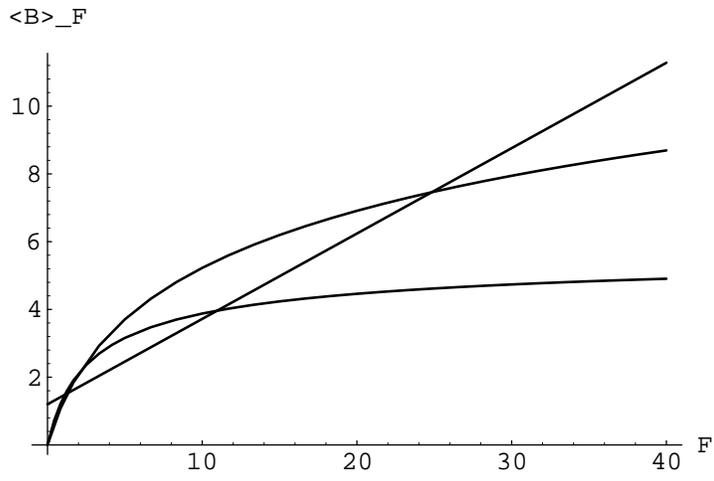
\end{document}